\documentclass[aps,pra,amsmath,amssymb,reprint,twocolumn]{revtex4-1}

\usepackage{graphicx}
\usepackage{dcolumn}
\usepackage{bm}
\usepackage{epstopdf}

\begin{document}

\preprint{}

\title{Measuring the gravitational acceleration with precision matter-wave velocimetry}

\author{G. D'Amico$^1$}
\author{L. Cacciapuoti$^2$}
\author{M. Jain$^1$}
\author{S. Zhan$^1$}
\email[]{Also at International Centre for Theoretical Physics (ICPT), Trieste, Italy.}
\author{G. Rosi$^1$}
\email[]{rosi@fi.infn.it}
\affiliation{$^1$INFN Sezione di Firenze, Dipartimento di Fisica e Astronomia \& LENS, Universit\`a di Firenze, via Sansone 1, I-50019 Sesto Fiorentino (FI), Italy \\
$^2$European Space Agency, Keplerlaan 1, 2200 AG Noordwijk, The Netherlands \\}

\date{\today}

\begin{abstract}
One of the major limitations of atomic gravimeters is represented by the vibration noise of the measurement platform, which cannot be distinguished from the relevant acceleration signal. We demonstrate a new method to perform an atom interferometry measurement of the gravitational acceleration without any need for a vibration isolation system or post-corrections based on seismometer data monitoring the residual accelerations at the sensor head. With two subsequent Ramsey interferometers, we measure the velocity variation of freely falling cold atom samples, thus determining the gravitational acceleration experienced by them. Our instrument has a fractional stability of $ 9 \times 10^{-6}$ at 1~s of integration time, one order of magnitude better than a standard Mach-Zehnder interferometer when operated without any vibration isolation or applied post-correction. Using this technique, we measure the gravitational acceleration in our laboratory, which is found in good agreement with a previous determination obtained with a FG5 mechanical gravimeter.
\end{abstract}


\maketitle 

Cold atom interferometry~\cite{Varenna2014} is today the state of the art for precision measurements of accelerations~\cite{Kasevich1992,Peters1999,Mueller2008,Gouet2008}, rotations~\cite{Gustavson1997,Gustavson2000,Canuel2006,Gauguet2009}, gravity gradient~\cite{Snadden1998,Guirk2002,Sorrentino2014,Duan2014,Santos2015,Wang2016,Damico2017}, and curvature~\cite{Rosi2015,Asenbaum2017}. The performance levels achieved so far find important applications, not only in fundamental physics, where atom-based tests of the Einstein's equivalence principle are flourishing~\cite{Tarallo2014,Schlippert2014,Zhou2015,Duan2016,Rosi2017}, but also in other areas of research such as geodesy, Earth observation, and field prospecting~\cite{DeAngelis2009}.

However, the sensitivity of absolute gravimeters is often limited by the seismic noise along the measurement axis that, as a consequence of the equivalence principle, cannot be distinguished from the gravity itself. Two different measurement strategies are currently used to solve this issue. The first relies on complex seismic isolation systems to reduce the acceleration noise on the instrument platform. The second combines the atomic sensor with a seismometer that performs a coarse acceleration measurement and allows to operate the atom interferometer in the fine measurement regime~\cite{Merlet2009,Geiger2011}. In the first case, the use of bulky seismic isolation platforms represents a major limitation to the development of compact and ruggedized instruments expected to operate in harsh environments and in the presence of high vibration noise. In the second case, the correction calculated from the mechanical accelerometer data and applied to the atomic transition probability introduces errors depending on the seismometer response. Below a few tens of Hz, mechanical accelerometers behave as low-pass filters thus reducing the rejection ratio for low frequency vibration noise and limiting the instrument performance.

In this paper, we demonstrate a new atom interferometry scheme to measure the gravitational acceleration, particularly suitable for measurements under severe conditions of vibration noise. We use our $^{87}$Rb atom interferometer to perform high-sensitivity velocimetry on freely falling cold atom samples prepared in a narrow velocity distribution~\cite{Carey2017,Carey2018}. Measuring velocities rather than accelerations has an obvious advantage: it attenuates the impact of high frequency seismic noise by a factor $1/(2\pi f)$, allowing a very good resolution without any need for seismic isolation or post-corrections from a mechanical accelerometer.

Standard atomic gravimeters are operated in a vertical Mach-Zehnder geometry, in which a $\pi/2-\pi-\pi/2$ pulse sequence splits the atomic wavefunctions, redirects, and finally recombines them at the output ports of the interferometer. The resulting phase measurement $\phi$ is proportional to the gravitational acceleration $g$. In our instrument, we interrogate freely falling atoms with a simpler Ramsey interferometer along the vertical direction. The pulse sequence is now composed of two $\pi/2$ Raman pulses, determining a phase shift of the atom interference fringes that, to leading order, is equal to
\begin{equation}
\phi = [k_{\text{eff}}(v + v_r/2) - \omega_D]T.
\label{eq_phase_shift}
\end{equation}
Here, $v$ is the wavepacket velocity, $T$ the time between the two interferometer pulses, $v_r$ the recoil velocity, and $\omega_D = \omega_{\text{eff}} - \omega_{\text{HFS}}$ accounts for the Doppler effect due to the vertical atomic motion; $\omega_{\text{eff}}$ and $\omega_{\text{HFS}}$ are the effective frequency of the Raman lasers and the resonance frequency of the ground state hyperfine doublet. We interrogate the freely falling atomic samples at two different times, $t_1$ and $t_2$, and we measure the differential phase $\Delta\phi$ from the atom interference fringes. From Eq.~\ref{eq_phase_shift},
\begin{align}
\Delta\phi &= [k_{\text{eff}}(v_2 - v_1) - (\omega_{D,2} - \omega_{D,1})]T \notag \\
&=(k_{\text{eff}}g - \alpha)TT_c,
\label{eq_differential_phase}
\end{align}
where $T_c = t_2 - t_1$ is the free-fall time between the two interferometer sequences and $\alpha = (\omega_{D,2} - \omega_{D,1})/T_c$ is the slope of the Doppler frequency ramp applied to the atoms to keep them in resonance with the Raman transition. From the velocity variation, which we measure interferometrically in an atom velocimetry experiment, we can determine the gravitational acceleration experienced by the atoms during their free fall.

\begin{figure}[t!]
\includegraphics[width=0.45\textwidth]{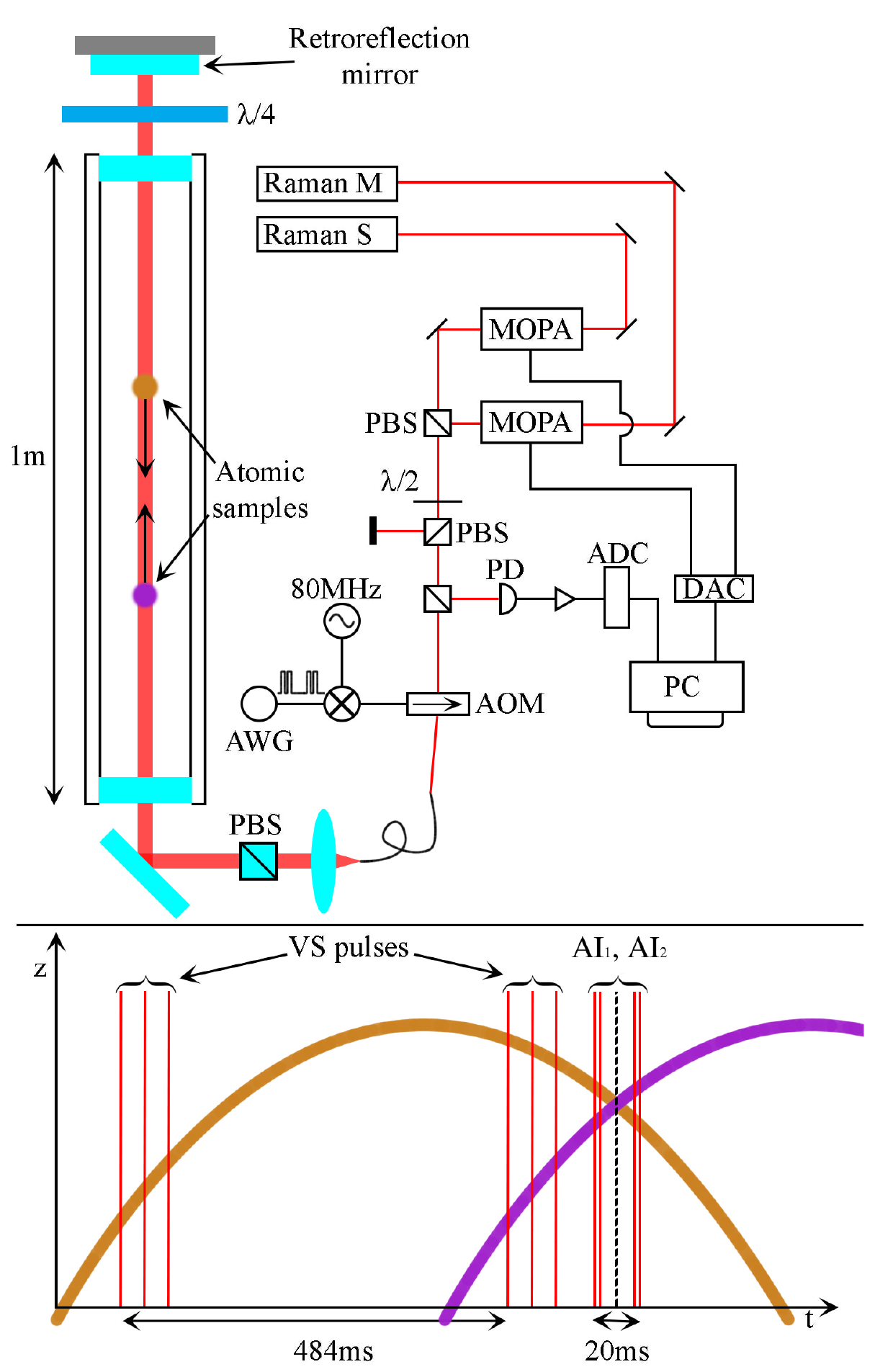}
\caption{(top) Scheme of the experimental apparatus. Two clouds of cold $^{87}$Rb atoms are launched vertically inside a vacuum tube. During the interferometer, they interact with counter-propagating Raman lasers derived from two independent MOPA oscillators. Their power output is stabilized by a digital proportional-integral loop, which steers their current with a $3$~Hz bandwidth. The two lasers are mixed in a polarizing beam splitter (PBS) and their polarizations are filtered and aligned in a second PBS. The velocity selection pulses and the interferometer pulses are generated by driving the acousto-optic modulator (AOM) with an arbitrary waveform generator (AWG). (bottom) Parabolic trajectories for the two atomic clouds (in orange and purple) and time sequence of the Raman pulses (in red) during the measurement cycle. The two atom interferometers AI$_1$ and AI$_2$ are taking place sequentially, when the two atomic samples are propagating with the same velocity, but in opposite directions. \label{figura_apparato}}
\end{figure}

Our experimental set-up is described in detail in~\cite{Sorrentino2014,Rosi2015}. In this article, we focus on the measurement principle. A cloud of $^{87}$Rb atoms is trapped in a three-dimensional magneto-optical trap and cooled down to a temperature of $\approx 4$~$\mu$~K (see Fig.~\ref{figura_apparato}~(top)).
The cloud is launched along the vertical direction inside a $1$~m long magnetically shielded vacuum tube. Just before entering the tube, we prepare the atomic sample in a narrow vertical velocity class with a corresponding temperature $T_z \approx 1$~nK and transfer them in the $(F = 1, m_F = 0)$ hyperfine level of the ground state with three consecutive counter-propagating Raman pulses. After $484$~ms, while the first sample is in free fall inside the interferometer tube, the same procedure is repeated to launch a second atomic cloud. With this sequence, we obtain two freely falling samples that are probed in two successive interferometric sequences (see Fig.~\ref{figura_apparato}~(bottom)), while they are moving with the same velocity, but with opposite directions: the atomic cloud launched last is interrogated at $t_1$ while it is propagating upwards (AI$_1$); viceversa, the atomic cloud launched first is interrogated at $t_2$ while it is propagating downwards (AI$_2$).

During the Ramsey interferometer, the atoms are interrogated on a $\pi/2-\pi/2$ pulse sequence. The lasers are resonant with the $6.8$~GHz Raman transition between the two hyperfine levels of the $^{87}$Rb ground state and have an effective wave vector $k_{\text{eff}} = 16\times 10^6$~m$^{-1}$; the master laser has a $2.2542$~GHz red detuning with respect to the $5^2S_{1/2}|F = 2\rangle \rightarrow 5^2P_{3/2}|F = 3\rangle$ transition. Aligned along the vertical axis, the lasers enter the vacuum system from the bottom and are retro-reflected by a mirror placed on top of the interferometer tube. In our setup, the retro-reflecting mirror is not seismically isolated. The Raman lasers are derived from two independent 1~W MOPA oscillators, whose power is stabilized by steering the laser current (see Fig.~\ref{figura_apparato}~(top)). The power stabilization loop is operated by a digital proportional-integral controller acting independently on the two MOPA oscillators with 3~Hz bandwidth. In this way, we control the intensity ratio between the slave and the master lasers to $I_\textrm{s} / I_\textrm{m} = 0.47$, thus cancelling systematics arising from the differential light shift. The velocity selection pulses and the interferometer pulses are shaped with an acousto-optical modulator driven by an arbitrary waveform generator. The two Raman beams reach the atoms via a common polarization-maintaining fibre; their polarization is cleaned by a Glenn-Taylor polarizer just before entering the vacuum system.  The time separation between the two $\pi/2$ pulses is limited to $T = 200$~$\mu$s; each pulse has a square envelope and a duration of $\tau = 12$~$\mu$s. Indeed, for $T>200$~$\mu$s, the separation between the atomic wavefunctions at the output ports of the interferometer becomes larger than the coherence length of the atoms, consequently reducing the fringes contrast. Due to the short interrogation time $T$, seismic noise is efficiently suppressed even if we do not apply any vibration isolation on the retro-reflecting mirror. In each experimental cycle, the atom interferometer pulse sequence $\pi/2-\pi/2$ is applied twice, at AI$_1$ and AI$_2$, separated by a time interval of $20$~ms. Considering that the second atomic cloud is launched $484$~ms after the first, the effective time separation between AI$_1$ and AI$_2$ is $T_c = 504$ ms. Due to the different velocities and Doppler shifts of the atoms, different frequency detunings need to be applied to the Raman lasers during the interferometers AI$_1$ and AI$_2$. In this way, AI$_1$ and AI$_2$ only act on the ascending and descending cloud, respectively. Fig.~\ref{figura_apparato}~(bottom) shows the atomic trajectories together with the time of the Raman pulses for the triple velocity selection and for the Ramsey interferometer. At the end of the sequence, we detect the laser-induced fluorescence emission from the two hyperfine levels of the $^{87}$Rb ground state and measure the normalized atomic population. The complete measurement cycle takes $\approx 1.9$~s. In order to efficiently reject systematic effects that do not depend on $k_{\text{eff}}$, for each measurement we apply the k-reversal procedure~\cite{Pereira2011}.

\begin{figure}[t]
\includegraphics[width=0.45\textwidth]{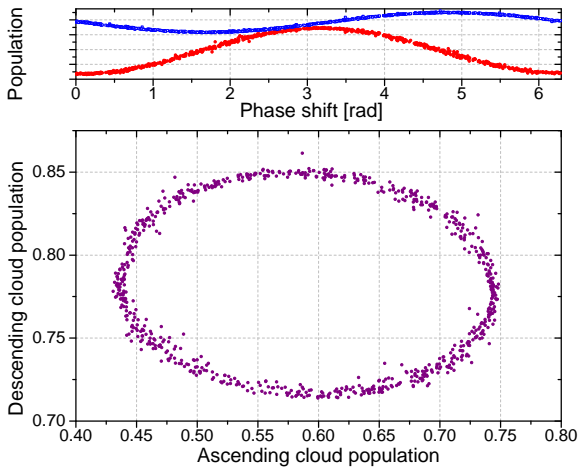}
\caption{(top) Fringes from the interferometer on the ascending (red circles) and descending (blue squares) clouds. (bottom) Lissajoux figure obtained by plotting the signal from the interferometer on the descending cloud as a function of the one on the ascending cloud. The relative phase difference $\Delta\phi$ of Eq.~\ref{eq_differential_phase} is obtained as a result of an elliptical fit.\label{fig_fringes_ellipse}}
\end{figure}

Figure~\ref{fig_fringes_ellipse}~(top) shows the atomic fringes from the interferometers AI$_1$ (red dots) and AI$_2$ (blue squares), acting on the ascending and descending clouds respectively.
The fringes are scanned by adding a common frequency detuning $\delta$ to the Doppler compensating frequencies for AI$_1$ ($\omega_{D,1}$) and AI$_2$ ($\omega_{D,2}$). This corresponds to a translation of the effective Doppler frequency ramp, without changing its slope $\alpha$ (see Eq.~\ref{eq_differential_phase}). The common frequency detuning $\delta$ is randomly chosen at each experiment cycle from a uniform distribution on the $0$ to $5$~kHz interval, covering a $2\pi$ phase for $T = 200$~$\mu$s. The slope $\alpha$ is set to a value that provides a relative phase difference between the two AI fringes close to $\pi/2$.  Figure~\ref{fig_fringes_ellipse}~(bottom) shows a typical Lissajous curve obtained by plotting the interference fringes on the descending cloud as a function of the ascending cloud fringes. From an elliptical fit, we retrieve the phase difference $\Delta\phi$. As expected, the signal from the interferometer on the descending cloud shows a lower contrast. Indeed, this cloud expands for $504$~ms longer than the ascending cloud before being interrogated, thus suffering from stronger heating effects.

Due to the finite duration $\tau$ of the interferometer pulses, Eq.~\ref{eq_phase_shift} is multiplied by an overall scale factor $A(\Omega_\textrm{eff})$, which depends on the effective Rabi frequency $\Omega_\textrm{eff}$
\footnote{
The scale factor $A(\Omega_\textrm{eff})$ has been calculated by using the sensitivity function approach~\cite{Dick1987}:
$A(\Omega_\textrm{eff}) = \frac{1}{\Omega_\textrm{eff}}[\sin^3\Omega_\textrm{eff}\tau(2-2\cos\Omega_\textrm{eff}\tau+\Omega_\textrm{eff}T\sin\Omega_\textrm{eff}\tau)]$
\label{A}
}.
This contribution, negligible in the limit $\tau \ll T$, plays a non negligible role in our setup, where $T = 200$~$\mu$s and $\tau = 12$~$\mu$s. Therefore, to extract the gravitational acceleration we exploit the linear dependence of $\Delta\phi$ from the slope $\alpha$, which can be easily changed in our experimental apparatus. We measure the phase difference $\Delta\phi$ for two different values of the slope $\alpha_{1,2}$ after applying the k-reversal technique. By linear interpolation, we determine the slope $\alpha_0 = k_{\text{eff}}g$ for which $\Delta\phi = 0$ and we extract the gravitational acceleration as $g=\alpha_0/k_{\text{eff}}$. After a 2-hour measurement run, we obtain $g = 9.8049234(21)$~ms$^{-2}$, consistent within $2\sigma$ with the previous measurement performed in our laboratory with the FG5 mechanical gravimeter, $g_\text{FG5} = 9.80492048(3)$~ms$^{-2}$~\cite{Angelis2012}.

\begin{figure}[t]
\includegraphics[width=0.45\textwidth]{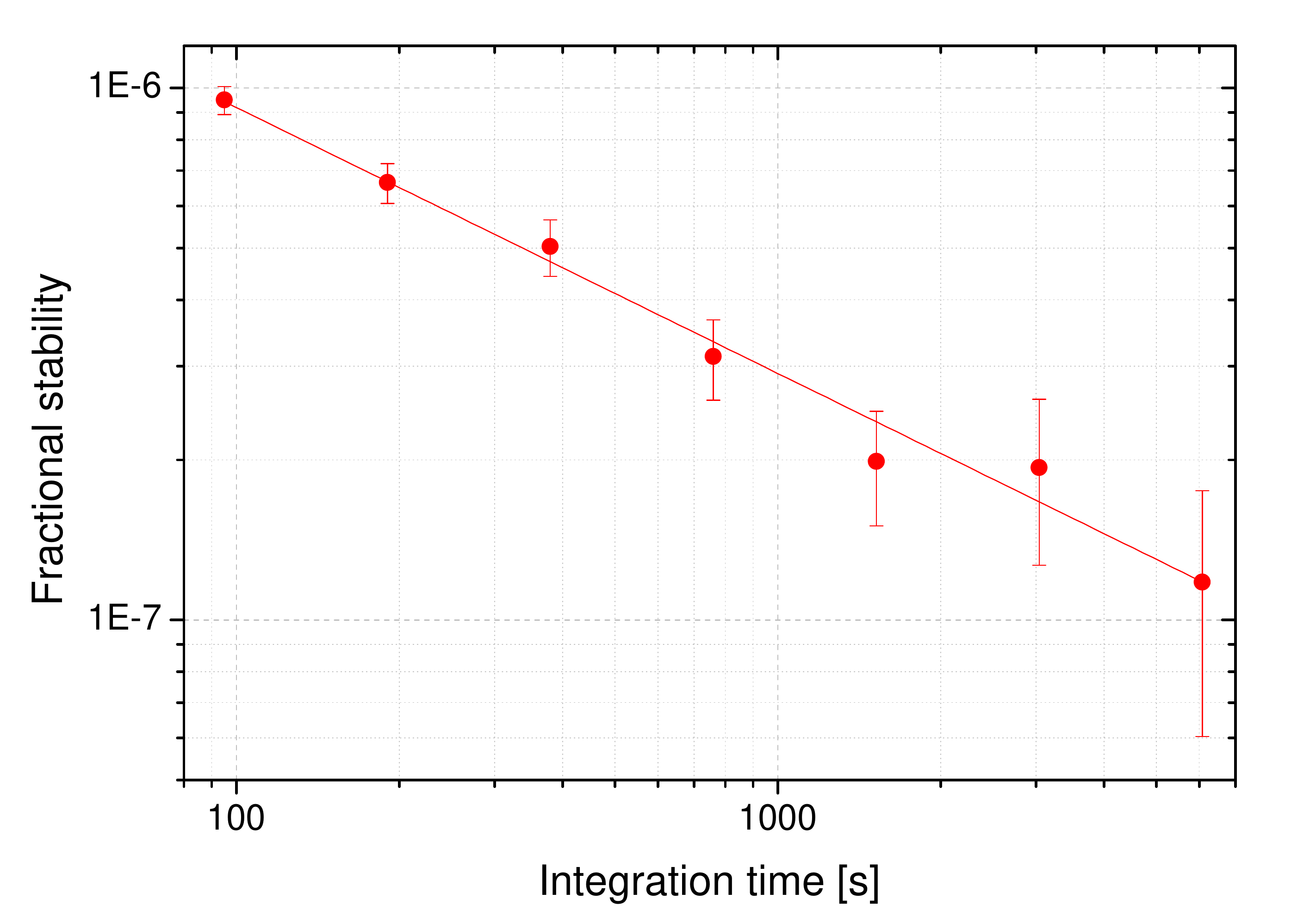}
\caption{Allan deviation of the acceleration measurements obtained after an 8-hour measurement run. The instrument has a fractional acceleration stability of $ 9 \times 10^{-6}$ at 1~s of integration time, averaging down to $1 \times 10^{-7}$ after $6000$ s. This performance is one order of magnitude better than what obtained with a standard Mach-Zehnder gravimeter without vibration isolation or post-correction of the seismic noise.\label{allan_measurement}}
\end{figure}

To evaluate the sensitivity of our gravimeter, we perform a 8-hour measurement campaign, during which we keep the same Doppler slope $\alpha$ and apply the k-reversal procedure. Figure~\ref{allan_measurement} shows the stability of the instrument evaluated in terms of Allan deviation.
The characteristic $1/\sqrt{t}$ slope of the curve shows that our measurements are dominated by white acceleration noise. The instrument has a fractional acceleration stability of $9 \times 10^{-6}$ at $1$~s of integration time, averaging down to $1 \times 10^{-7}$ after $6000$~s. The long term stability and accuracy of the measurement is limited by residual light shift effects, which introduce velocity variations during the preparation phase, when atoms are selected in a narrow velocity class before entering the Ramsey interferometer. From that point of view, the stabilization of the Raman laser powers is of crucial importance to control light shift effects.

\begin{figure}[t]
\includegraphics[width=0.45\textwidth]{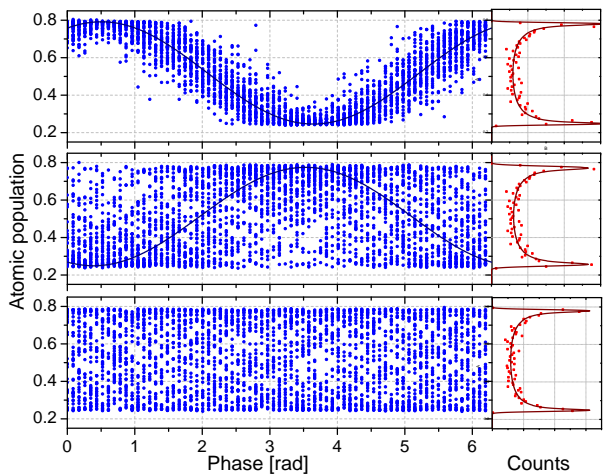}
\caption{Atom interference fringes measured in a standard Mach-Zehnder interferometer for different interrogation times, $T = 5,\,10,\,20$~ms (from top to bottom), without any seismic isolation or post-correction. Each data set corresponds to a $90$~min measurement run. We also show the corresponding sinusoidal fits. Due to the seismic noise, the sinusoidal fit for $T = 20$~ms does not converge to a stable solution. On the right hand panel, we report the histograms of the relative atomic populations together with the corresponding fits~\cite{Geiger2011}. In the best conditions, achieved for an interrogation time $T = 10$~ms, we evaluate a fractional stability for the acceleration measurements of $9 \times 10^{-5}$ at $1$~s of averaging time \label{fig_triple_fringe}}
\end{figure}

Finally, we compare the stability of this measurement method to the results of a standard Mach-Zehnder interferometer. To this purpose, we interrogate the atoms on a $\pi/2-\pi-\pi/2$ Raman pulse sequence and we measure the gravitational acceleration for a total duration of $90$~min. Figure~\ref{fig_triple_fringe} shows the atom interference fringes measured for three different interrogation times: $5$, $10$ and $20$~ms (from top to bottom). These measurements are performed in the same experimental setup as before; therefore, they are affected by the same levels of seismic noise.
On the right hand panel of Fig.~\ref{fig_triple_fringe}, we also plot the histograms of the relative populations and the corresponding fits~\cite{Geiger2011}. For these particular measurements, the fringes are scanned acting on the relative phase between the Raman lasers. Since the retro-reflecting mirror is not seismically isolated, atom interference fringes are quickly washed out, with the sinusoidal fit to the data failing to converge already for $T=20$~ms. The best measurement conditions are found for $T = 10$~ms, corresponding to a stability of $9 \times 10^{-5}$ at $1$~s of averaging time. In the absence of vibration isolation or post-correction from a seismometer monitoring the vibration noise of the platform, the standard Mach-Zehnder interferometer is outperformed by the simpler method based on atom velocimetry by about one order of magnitude.

In conclusion, we have demonstrated a new scheme to measure the gravitational acceleration in the presence of high seismic noise. Our method uses two $\pi/2-\pi/2$ Ramsey sequences to measure the velocity variation of freely falling samples of atoms. The short term stability of the instrument is limited by the interrogation time $T$ and the free-fall time $T_c$. The long term stability and accuracy depend on the velocity variations introduced by the light shift during the preparation phase. The free-fall time and the interrogation time are both strictly related to the temperature of the atomic sample. Therefore, ultracold atom sources becomes highly beneficial for several reasons: they increase the efficiency of the velocity selection process (down to the pK regime), while preserving high atom numbers; they allow longer free evolution times; they provide atomic samples with larger coherence lengths that can be probed in longer interferometers with limited loss of contrast. The free-fall time can also be extended by using coherent Bloch oscillations in a vertical optical lattice. Finally, light shift compensation techniques, as proposed in~\cite{Kovachy2015}, can be implemented to push further the measurement stability and accuracy. As an example, with $^{88}$Sr atoms in a vertical optical lattice, $T_c$ can be increased up to 10~s or more~\cite{Ferrari2006}. Moreover, the ultra-narrow Sr optical clock transition can be used to realize a velocity selection with nearly zero light shift~\cite{Hu2017}, thus improving both the long term stability and the accuracy of the measurement. From that point of view, our method shows very interesting perspectives for being further developed towards state-of-the-art performance and beyond.

\begin{acknowledgments}
The authors acknowledge stimulating discussions with G.M. Tino. This work was supported by INFN (MAGIA advanced experiment).
\end{acknowledgments}

\bibliography{biblio}

\end{document}